\documentclass[twocolumn]{aastex631}

\usepackage{graphicx}
\usepackage{mathptmx}
\usepackage{bm}
\usepackage{esvect}
\usepackage{amsmath}
\usepackage{multirow}
\usepackage{soul}

\graphicspath{{./}}

\begin{document}

\title{PSR J1231$–$1411 revisited: Pulse Profile Analysis of X-ray Observation}

\correspondingauthor{Mingyu GE}
\email{gemy@ihep.ac.cn}
\correspondingauthor{Shuang-Nan ZHANG}
\email{zhangsn@ihep.ac.cn}
\correspondingauthor{Fangjun LU}
\email{lufj@ihep.ac.cn}

\author{Liqiang QI}
\affiliation{Key Laboratory of Particle Astrophysics, Institute of High Energy Physics, Chinese Academy of Sciences, Beijing 100049, China}
\author{Shijie ZHENG}
\affiliation{Key Laboratory of Particle Astrophysics, Institute of High Energy Physics, Chinese Academy of Sciences, Beijing 100049, China}
\author{Juan ZHANG}
\affiliation{Key Laboratory of Particle Astrophysics, Institute of High Energy Physics, Chinese Academy of Sciences, Beijing 100049, China}
\author{Mingyu GE}
\affiliation{Key Laboratory of Particle Astrophysics, Institute of High Energy Physics, Chinese Academy of Sciences, Beijing 100049, China}
\author{Ang LI}
\affiliation{Department of Astronomy, Xiamen University, Xiamen 361005, China}
\author{Shuang-Nan ZHANG}
\affiliation{Key Laboratory of Particle Astrophysics, Institute of High Energy Physics, Chinese Academy of Sciences, Beijing 100049, China}
\affiliation{University of Chinese Academy of Sciences, Chinese Academy of Sciences, Beijing 100049, China}
\author{Fangjun LU}
\affiliation{Key Laboratory of Particle Astrophysics, Institute of High Energy Physics, Chinese Academy of Sciences, Beijing 100049, China}
\author{Hanlong PENG}
\affiliation{Key Laboratory of Particle Astrophysics, Institute of High Energy Physics, Chinese Academy of Sciences, Beijing 100049, China}
\author{Liang ZHANG}
\affiliation{Key Laboratory of Particle Astrophysics, Institute of High Energy Physics, Chinese Academy of Sciences, Beijing 100049, China}
\author{Hua FENG}
\affiliation{Key Laboratory of Particle Astrophysics, Institute of High Energy Physics, Chinese Academy of Sciences, Beijing 100049, China}
\author{Zhen Zhang}
\affiliation{Key Laboratory of Particle Astrophysics, Institute of High Energy Physics, Chinese Academy of Sciences, Beijing 100049, China}
\author{Yupeng XU}
\affiliation{Key Laboratory of Particle Astrophysics, Institute of High Energy Physics, Chinese Academy of Sciences, Beijing 100049, China}
\affiliation{University of Chinese Academy of Sciences, Chinese Academy of Sciences, Beijing 100049, China}
\author{Zhengwei LI}
\affiliation{Key Laboratory of Particle Astrophysics, Institute of High Energy Physics, Chinese Academy of Sciences, Beijing 100049, China}
\author{Liming Song}
\affiliation{Key Laboratory of Particle Astrophysics, Institute of High Energy Physics, Chinese Academy of Sciences, Beijing 100049, China}
\author{Shu ZHANG}
\affiliation{Key Laboratory of Particle Astrophysics, Institute of High Energy Physics, Chinese Academy of Sciences, Beijing 100049, China}
\author{Lian TAO}
\affiliation{Key Laboratory of Particle Astrophysics, Institute of High Energy Physics, Chinese Academy of Sciences, Beijing 100049, China}
\affiliation{University of Chinese Academy of Sciences, Chinese Academy of Sciences, Beijing 100049, China}
\author{Wentao YE}
\affiliation{Key Laboratory of Particle Astrophysics, Institute of High Energy Physics, Chinese Academy of Sciences, Beijing 100049, China}

\begin{abstract}
One of the primary goals of Neutron Star Interior Composition Explorer (NICER)-like X-ray missions is to impose stringent constraints on the neutron star equation of state by precisely measuring their masses and radii. NICER has recently expanded the dataset of inferred mass-radius relations for neutron stars, including four rotation-powered millisecond pulsars PSR J0030+0451, PSR J0740+6620, PSR J0437--4715, and PSR J1231--1411. In this work, the mass--radius relation and X-ray emitting region properties of PSR J1231--1411 are inferred with an independent pulse profile modeling based on the spherical star Schwarzschild-spacetime and Doppler approximation. With one single-temperature elongated hot spot and one single-temperature crescent hot spot, the inferred gravitational mass is $M~=~1.12~\pm~0.07$~\(\textup{M}_\odot\) and the inferred equatorial radius is $R_\textup{eq}~=~9.91_{-0.86}^{+0.88}$~km (68\% credible intervals). It provides an alternative geometry configuration of the X-ray emitting region for PSR J1231--1411 to sufficiently explain the observation data of NICER and XMM-Newton. The inferred radius is smaller than that derived by \citet{salmi2024nicer} ($M~=~1.04_{-0.03}^{+0.05}$~\(\textup{M}_\odot\), $R_\textup{eq}~=~12.6~\pm~0.3$~km), and the inferred mass is slightly higher in this work. The inferred geometry configurations of the X-ray emitting region in both works are non-antipodal, which is not consistent with a centered dipole magnetic field and suggests a complex magnetic field structure. 
\end{abstract}


\section{Introduction}
\label{sec1}
The mass ($M$) and radius ($R$) measurements of neutron stars probe the supranuclear density matter within their interiors and provide constraints on the equation of state~\citep{lattimer2001neutron}. Many methods are available to constrain the equation of state through radio, X-ray, and gravitational wave observations~\citep{yunes2022gravitational}. 
Measuring pulsed thermal emission from hot regions on a rotating neutron star~\citep{gendreau2016neutron} is considered one of the most promising techniques for providing accurate and precise constraints on $M/R$~\citep{2024APh...15802935A}. Recent studies using the Neutron Star Interior Composition Explorer (NICER) data on nearby rotation-powered millisecond pulsars, including PSR J0030+0451, PSR J0740+6620, PSR J0437--4715, and PSR J1231--1411, have advanced the understanding of the neutron star structure via pulse profile modeling.
NICER fits the observed energy-resolved pulse profiles and allows the Bayesian parameter inference of the mass--radius relation and X-ray emitting region properties~\citep{miller2019psr,riley2019nicer,salmi2023atmospheric,vinciguerra2024updated,miller2021radius,riley2021nicer,salmi2022radius,salmi2024radius,dittmann2024more,choudhury2024nicer,salmi2024nicer}. These inferred parameters are helpful in constraining the currently conflicting equation of state~\citep[e.g.][]{watts2016colloquium,miller2019psr,raaijmakers2021constraints,miao2024thermal} and providing hints on the magnetic field structure and X-ray emission mechanism~\citep{bilous2019nicer,chen2020numerical,kalapotharakos2021multipolar,carrasco2023relativistic}. 

The theoretical framework of the pulse profile modeling is based on the oblate star Schwarzschild-spacetime and Doppler approximation to track photons from the stellar surface to the X-ray telescope through the exterior spacetime of rotating neutron stars~\citep[e.g.][]{poutanen2003nature,poutanen2006pulse,cadeau2007light,morsink2007oblate,algendy2014universality,nattila2018radiation,bogdanov2019constrainingii}. The radiative model of a geometrically thin atmosphere is also incorporated to modify the energy spectrum and anisotropic distribution of the thermal emission from the stellar surface~\citep{ho2001atmospheres,ho2009neutron}. In recent studies, the fully ionized and non-magnetic NSX hydrogen atmosphere has been shown to describe the observation data adequately~\citep{salmi2023atmospheric,choudhury2024nicer,salmi2024nicer}. The numerical algorithms following this theoretical framework are implemented~\citep{miller2019psr,riley2019nicer,bogdanov2021constraining}, e.g.\ the X-ray Pulse Simulation and Inference (X-PSI) software package~\citep{riley2023x} and another independent pipeline presented in \cite{miller2019psr,miller2021radius} and \cite{dittmann2024more}. The Bayesian analysis of these four pulsars is mainly carried out by the two software packages. 

The measurement precision available from recent NICER studies is hindered by features specific to each of these sources. PSR J0030+0451 is an isolated pulsar with no prior information on the mass and view inclination, which makes it challenging to perform sufficient and proper sampling in a large model parameter space~\citep{miller2019psr,riley2019nicer,salmi2023atmospheric,vinciguerra2024updated}. The importance and necessity of the mass and inclination prior information in the pulse profile modeling have been reported in \cite{choudhury2024nicer}. PSR J0740+6620 is a binary pulsar with well-constrained mass and inclination prior information derived from the radio timing observations~\citep{miller2021radius,riley2021nicer,salmi2022radius,salmi2023atmospheric,salmi2024radius,dittmann2024more}. However, it is very faint, with the majority of the total NICER energy spectrum being the non-source background. The inferred equatorial radius is still in a wide distribution with the currently available photon statistics. PSR J0437--4715 is a binary pulsar with well-constrained mass, inclination, and distance prior information~\citep{choudhury2024nicer}. However, a bright Seyfert II Active Galactic Nuclei (AGN) is present in the NICER field of view, which provides a non-negligible contribution to the non-source background. The NICER background model cannot eliminate source contamination inside the field of view, whose treatment affects the inferred parameters with the background-marginalized likelihood function. PSR J1231--1411 is a binary pulsar, but the preliminary radio timing measurement provides much less tight constraints on the mass and inclination~\citep{salmi2024nicer}. The distance has multiple solutions from independent measurements. Additionally, the weak inter-pulse in the pulse profile complicates the sampling process in the model parameter space, as noted in \citet{salmi2024nicer}. 

In this work, PSR J1231--1411 is re-analyzed based on the previous study~\citep{salmi2024nicer} using an independent pulse profile modeling with the spherical star Schwarzschild-spacetime and Doppler approximation. The structure of the paper is organized as follows: the methodology of the pulse profile modeling is briefly reviewed in Section~\ref{sec3}, as well as the code validation; the processing of PSR J1231--1411 observations from NICER and XMM-Newton is presented in Section~\ref{sec2}; its inferred model parameters are discussed in Section~\ref{sec4}; finally, the conclusions are given in Section~\ref{sec5}. 

\section{Methodology}
\label{sec3}
\subsection{Pulse profile modeling} 
An independent numerical algorithm written in C++ is implemented in this work with the spherical star Schwarzschild-spacetime and Doppler approximation~\citep[e.g.][]{poutanen2003nature,poutanen2006pulse,cadeau2007light,morsink2007oblate,algendy2014universality}. The neutron star is squeezed into an oblate spheroid as the rotation speed increases, which plays a vital role in the pulse profile modeling for rapidly rotating neutron stars~\citep{morsink2007oblate}. However, regarding the millisecond pulsars studied in this work with the spin frequency being below $\approx$~300~Hz, the effect of oblateness is insignificant on the parameter inference of the gravitational mass and equatorial radius~\citep{cadeau2007light,morsink2007oblate}. Though the theoretical framework of the pulse profile modeling is well established concerning the relativistic effects due to the strong gravitational field and rapid spin~\citep[e.g.][]{bogdanov2019constrainingii}, the crucial steps for implementing the numerical algorithm are reviewed and presented in this section for clarity. 

The light-bending and time-delay effects are directly obtained by calculating the lensing integral instead of numerically solving the differential equations of motion in the ray tracing~\citep{cadeau2007light,morsink2007oblate}. According to special relativity, quantities are converted from the comoving frame to the static frame, and the radiation from an area element on the stellar surface is obtained accordingly.

The thermal emission of hot regions is modified by the existence of a centimeter-thick atmosphere surrounding neutron stars, which can be calculated by detailed radiative models, e.g.\ the fully ionized and non-magnetic NSX hydrogen atmosphere model by \cite{ho2001atmospheres} and \cite{ho2009neutron}. It has been widely used in previous studies and has been proven to reproduce the observation dataset adequately~\citep{salmi2023atmospheric,choudhury2024nicer,salmi2024nicer}. The atmosphere model is provided (nsx\_H\_v200804 is used in this work~\citep{nsxmodel}) as a pre-computed lookup table with four independent variables to perform the interpolation of the local radiation intensity as a function of the effective temperature, effective gravity, photon energy, and cosine of the emission angle. 

The observed flux $F_{\Phi_k,E_j}$ on the image plane at the rotation phase $\Phi_k$ and energy $E_j$, i.e.\ in each phase-energy bin, can be obtained by numerically integrating the radiation contribution from all discretized meshes of the hot regions. It then multiplies an attenuation factor $\exp(-N_\textup{H} \sigma_\textup{abs})$ to account for the interstellar absorption. The absorption cross-section $\sigma_\textup{abs}$ can be obtained from a lookup table of the TBabs model~\citep{wilms2000absorption}. Finally, the registered flux $S_{\Phi_k, E_i}$ measured by the telescope is obtained by convoluting the flux $F_{\Phi_k}$ with the instrument response $R_{ij}$, where the first index $i$ denotes the data channel and the second index $j$ denotes the model energy bin number.

\subsection{Parameter estimation and model evaluation}
Bayesian analysis is usually more efficient and robust in a high-dimensional problem compared to the maximum likelihood estimation. Since the number of free parameters typically ranges from 15 to 30 in the pulse profile modeling, Bayesian analysis is more suitable to fit the observation dataset and test the model. The widely used Bayesian techniques in astrophysics include Markov Chain Monte Carlo (MCMC)~\citep{foreman2013emcee} and Nested Sampling Monte Carlo (NSMC)~\citep{skilling2004nested}. Though a hybrid approach combining NSMC and MCMC methods was proposed in \cite{miller2019psr}, only the NSMC method is used in this work due to limited computation resources. The publicly available nested sampler MultiNest~\citep{feroz2009multinest} is adopted. 

\subsection{Validation}
The accuracy of the numerical algorithm implemented in this work is validated by re-analyzing the NICER observations of PSR J0030+0451, PSR J0740+6620, and PSR J0437--4715. Two important aspects need to be clarified, including the hot spot geometry and non-source background. First, the properties of the X-ray emitting regions, i.e.\ the shape and temperature distribution, remain uncertain. A series of simplified geometric models have been proposed motivated by the physics of magnetospheric return currents~\citep{harding2001pulsar}, including circles, rings, crescents, ovals, and their combinations with single or dual temperature distribution~\citep{riley2019nicer,miller2019psr}. In this work, the recommended geometric model for each source is selected for the validation, i.e.\ one single-temperature circular region and one single-temperature crescent region (\textbf{ST+PST}) for PSR J0030+0451~\citep{riley2019nicer}, two single-temperature circular regions (\textbf{ST-U}) for PSR J0740+6620~\citep{miller2021radius,riley2021nicer}, one single-temperature annulus region and one dual-temperature overlapping region (\textbf{CST+PDT}) for PSR J0437--4715~\citep{choudhury2024nicer}. 

Second, these sources are very faint, such that the contribution from the pulsar is comparable to or even less than the non-source background in the registered energy spectrum. The non-source background mainly includes the diffuse X-ray background, instrumental background, and nearby source contamination. To better infer the mass--radius relation and X-ray emitting region properties with reasonable computation speed, \cite{miller2019psr} and \cite{riley2019nicer} proposed the background-marginalized likelihood ,
\begin{equation}
\mathcal{L}_\textup{NICER}({D} \mid \boldsymbol{\theta}, \mathcal{M}) \propto \int_{B_\textup{l}}^{B_\textup{u}} \mathcal{L}_\textup{NICER}({D} \mid \boldsymbol{\theta}, \mathcal{M},B) dB \,,
\end{equation}
where ${D}$ is the observation data, $\boldsymbol{\theta}$ is the model parameter of a given model $\mathcal{M}$, and $B$ is the expected background counts. In each channel of the energy spectrum, the likelihoods corresponding to each possible number of backgrounds are numerically integrated. The possible number of backgrounds in each channel can be estimated based on the NICER background model with a lower limit $B_\textup{l}$ and an upper limit $B_\textup{u}$. The background limits can be important for parameter inference, especially in the case of PSR J0437--4715 with non-negligible source contamination in the field of view~\citep{choudhury2024nicer}. The Poisson likelihood function is used considering the limited number of counts in each phase-energy bin ($\Phi_\textup{i}$, $E_\textup{j}$), 
\begin{equation}
\begin{aligned}
\ln \mathcal{L}_\textup{NICER}({D} \mid \boldsymbol{\theta}, \mathcal{M},B) = & \sum_{\phi_\textup{i}} \sum_{E_\textup{j}} D_\textup{ij} \ln(M_\textup{ij}+B_\textup{ij}) \\ 
&- (M_\textup{ij}+B_\textup{ij}) - \ln |\Gamma(D_\textup{ij}+1)| \,,
\end{aligned}
\end{equation}
where $\Gamma$ is the gamma function. Additionally, the energy spectrum of XMM-Newton observations can be used to constrain the non-source background better because the focusing telescopes have a much higher source-to-background ratio. Thus, a joint fit is usually performed for the NICER and XMM-Newton observation dataset. The total log-likelihood, which takes the model parameters as input and returns its value to the sampler MultiNest, is defined as follows~\citep{miller2021radius,riley2021nicer},
\begin{equation}
\ln \mathcal{L}_\textup{total} = \ln \mathcal{L}_\textup{NICER} + \ln \mathcal{L}_\textup{XMM-Newton} \,,
\end{equation}
where $\mathcal{L}_\textup{XMM-Newton}$ can be EPIC-pn, EPIC-MOS, or multiple detectors.

\begin{figure}[h]
\begin{center}
\includegraphics[width=0.99\linewidth]{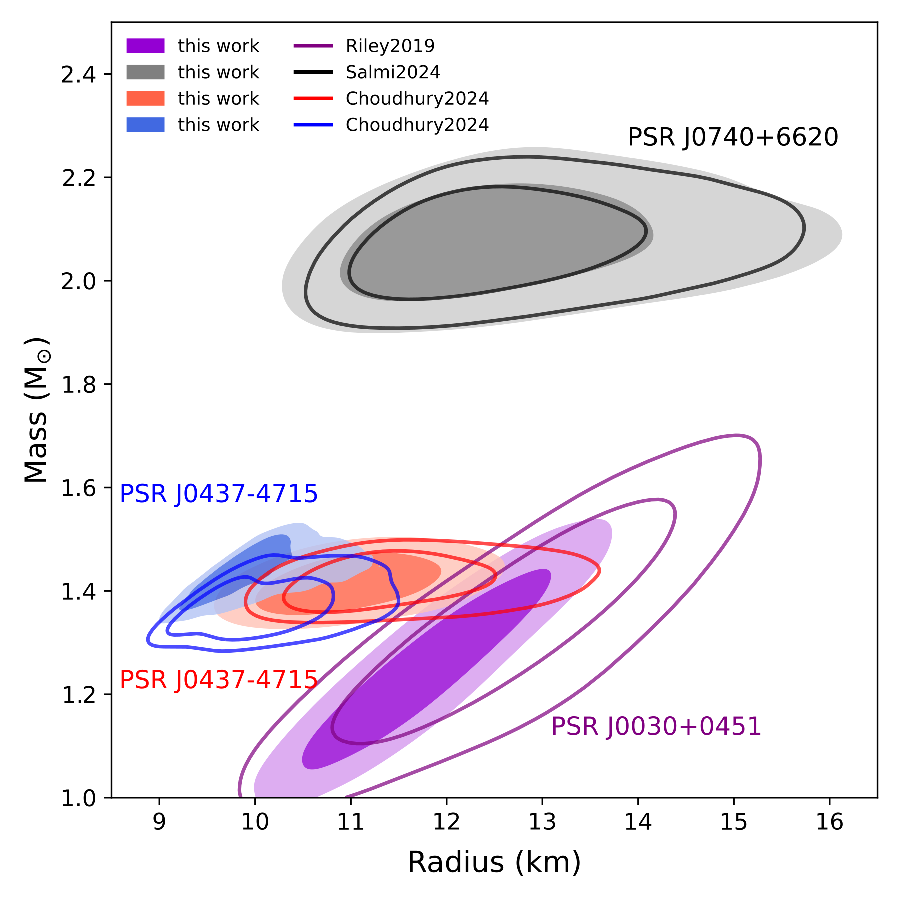}
\end{center}
\caption{2-D marginalized posteriors of masses and radii of three rotation-powered millisecond pulsars PSR J0030+0451, PSR J0740+6620, and PSR J0437--4715. The contours in the 2-D marginalized posterior denote the 68\% and 95\% credible intervals. The inferred result of \cite{riley2019nicer} for PSR J0030+0451 is plotted with purple lines, which uses the NICER-only data. The inferred result of \cite{salmi2024radius} for PSR J0740+6620 is plotted with black lines, which uses the NICER+XMM data. The inferred results of \cite{choudhury2024nicer} for PSR J0437--4715 are plotted with red lines and blue lines for the NICER-only and NICER-XMM data, respectively. In comparison, the inferred results using the spherical star Schwarzschild-spacetime and Doppler approximation in this work are plotted in shaded areas with different colors. The data is from \cite{J0030dataused}, \cite{J0740data}, and \cite{J0437data}, respectively. The numerical details are summarized in Table~\ref{validation_table}. } 
\label{validation_figure}
\end{figure}

\begin{table*}[ht]
\centering
\caption{Summary of the inferred and best-fit masses and radii of three rotation-powered millisecond pulsars PSR J0030+0451, PSR J0740+6620, and PSR J0437--4715 in this work.}
\begin{tabular}{lcccccccc}
\hline
		& \multicolumn{2}{c}{PSR J0030+0451} 			& \multicolumn{2}{c}{PSR J0740+6620}	& \multicolumn{2}{c}{PSR J0437--4715}    & \multicolumn{2}{c}{PSR J0437--4715}\\	
		& \multicolumn{2}{c}{NICER+XMM} 			    & \multicolumn{2}{c}{NICER+XMM}	        & \multicolumn{2}{c}{NICER only}         & \multicolumn{2}{c}{NICER+XMM}\\		        
Parameter	& $\widehat{\textup{CI}}_{68\%}$ & Best-fit	& $\widehat{\textup{CI}}_{68\%}$ & Best-fit   & $\widehat{\textup{CI}}_{68\%}$	& Best-fit  & $\widehat{\textup{CI}}_{68\%}$	& Best-fit\\
\hline
$F_0$ (Hz)		    & 205.53, fixed&-				& 346.53, fixed&-			& 173.69, fixed&-   & 173.69, fixed&-\\			
$M$ (\(\textup{M}_\odot\))  & 1.26$_{-0.13}^{+0.13}$ &1.39		& 2.07$_{-0.07}^{+0.07}$ &2.00		& 1.41$_{-0.04}^{+0.04}$ &1.46  & 1.44$_{-0.04}^{+0.04}$ &1.47\\
$R_\textup{eq}$ (km)	    & 11.82$_{-0.84}^{+0.84}$ &13.48 		& 12.66$_{-1.12}^{+1.15}$ &11.58 	& 11.00$_{-0.64}^{+0.66}$ &11.60  & 9.98$_{-0.41}^{+0.41}$ &10.18\\
\hline
\end{tabular}
\label{validation_table}
\end{table*}

Unless specified otherwise in this work, a joint fit to the NICER and XMM-Newton data is performed with the suggested geometric model for each source. The lower bound of the NICER background is set to zero, and the upper bound is set to the maximum possible number of counts in each channel. The background prior is uniformly distributed between the lower and upper bound. A single-background likelihood function is used for the XMM-Newton data due to a high source-to-background ratio. The main settings of MultiNest are the same for each source as follows: the sampling efficiency of 0.01, evidence tolerance of 0.1, live points of 1000, and the multi-modal option turned off. The only exception is the case of PSR J0437--4715, where the lower and upper bounds of the NICER background consider the 3C50 model and the AGN spectrum. The recommended model by \cite{choudhury2024nicer} for PSR J0437--4715 does not include the XMM-Newton constraints. The 2-D marginalized posteriors of masses and radii of three rotation-powered millisecond pulsars are plotted in Figure~\ref{validation_figure}. The corresponding inferred and best-fit parameters are summarized in Table~\ref{validation_table}. 

A good agreement can be seen in the case of PSR J0740+6620. Though the inferred masses and radii of PSR J0030+0451 and PSR J0437--4715 differ from previous analyses~\citep{miller2019psr,riley2019nicer,salmi2023atmospheric,vinciguerra2024updated,choudhury2024nicer}, the results are still consistent with their works. The possible reason for the discrepancies is the use of the spherical Schwarzschild-spacetime instead of the oblate Schwarzschild-spacetime~\citep{bogdanov2019constrainingii}. The inferred hot-spot geometries of PSR J0740+6620 are closer to the equator and thus less sensitive to this approximation. Additionally, the MultiNest settings differ from other works, which may be sensitive to the inferred parameters~\citep{vinciguerra2024updated}. A minimal sampling efficiency of 0.01 is used in this work instead of a large number of live points, because it decreases the computation time during the convergence test on the data of PSR J0740+6620. Though the proper MultiNest settings are usually problem-dependent, the settings are much finer than the recommended ones~\citep{feroz2009multinest} and are used for different celestial sources in this work. In addition, a single-background likelihood function is used for the XMM-Newton data instead of a background-marginalized likelihood function. Concerning the results of PSR J0030+0451, the solutions are discrepant and exhibit multi-modal structures. The solution of \cite{riley2019nicer} is used as the reference in Figure~\ref{validation_figure}, while the data of \cite{J0030dataused} is used for the Bayesian inference in this work. Thus, the observation data, instrumental response, model parameter space, and inclusion of XMM-Newton data can modify the inferred parameters. It should also be noted that it is very challenging to sample the model parameter space thoroughly in high-dimensional and multi-modal problems, especially in the case of PSR J0030+0451, where no prior information on the mass and inclination is available. Overall, the validation demonstrates a proper implementation of the numerical algorithm and the parameter inference process for pulsars with a spin frequency below $\approx$~300~Hz. 

\section{Observation data processing}
\label{sec2}
\subsection{NICER}
The NICER X-ray Timing Instrument (XTI) dataset~\citep{gendreau2016neutron} is used for the Bayesian analysis of the mass--radius relation and hot spot properties of PSR J1231--1411. The observation used in this work starts at 57930~MJD (ObsID 0060060101) and ends at 60512~MJD (ObsID 7060060734). The data processing follows the standard way suggested by the NICER group via \textit{nicerl2} (data analysis software HEAsoft version 6.33 and calibration database version CALDB xti20240206). The undershoot range is selected between 0 and 50 to minimize the effect of optical loading\footnote{https://heasarc.gsfc.nasa.gov/docs/nicer/analysis\_threads/light-leak-overview/} on the measured energy spectrum. The overshoot range is selected between 0 and 2 to reduce the impact of particle events on the measured energy spectrum~\citep{remillard2022empirical}. Based on \cite{bogdanov2019constrainingi}, a further data filtering process is performed on the cleaned data events to enhance the data quality, which excludes those with a counting rate (20-second bins) above 1.5 counts per second in the energy range between 0.3 and 1.5~keV. Consequently, the final effective exposure is reduced to 2.23~Ms, and the final spectrum contains 1156031 counts in the energy range between 0.3 and 1.5~keV. 

\begin{figure}[h]
\begin{center}
\begin{minipage}{0.48\linewidth}
\includegraphics[width=1.0\linewidth]{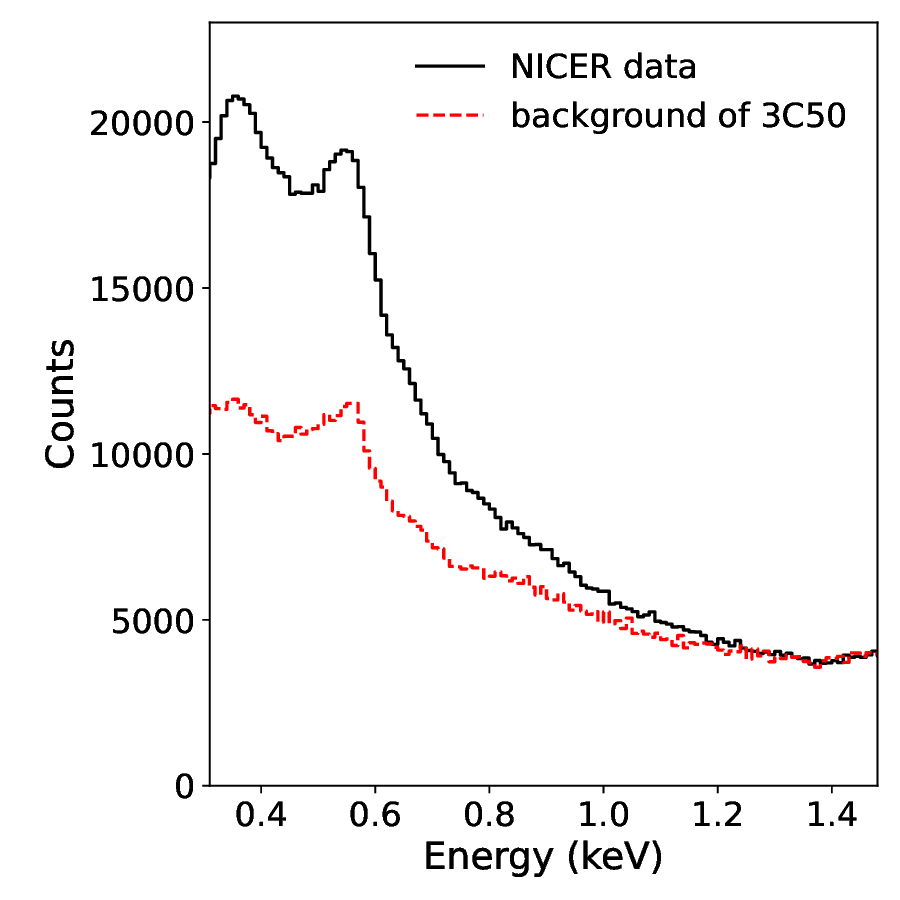}
\end{minipage}
\begin{minipage}{0.48\linewidth}
\includegraphics[width=1.0\linewidth]{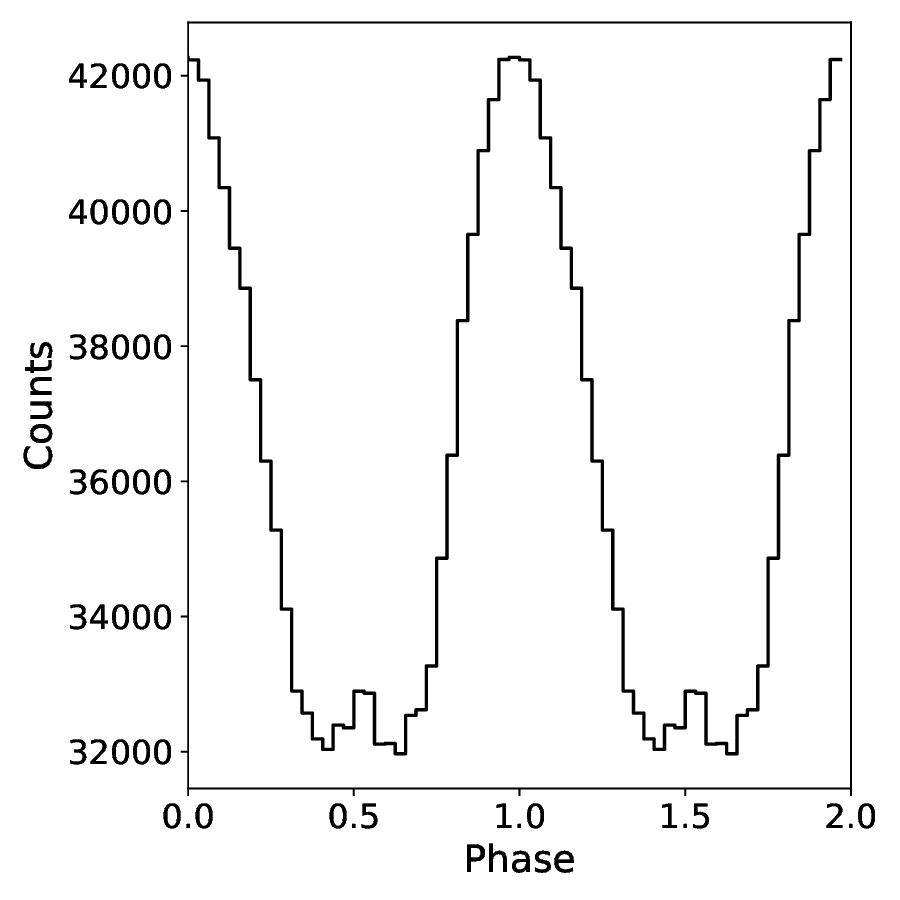}
\end{minipage}
\end{center}
\caption{Total NICER energy spectrum of PSR J1231--1411 with the background estimation from the 3C50 model (left panel). Energy-integrated pulse profile of PSR J1231--1411 in the range between 0.3 and 1.5~keV with 32 phase bins (right panel). Two rotational cycles are plotted for clarity.}
\label{NICER_observation}
\end{figure}

The background and instrumental responses are extracted by a single pipeline task \textit{nicerl3-spect}. Multiple models are available for the background estimation, including the SCORPEON, 3C50, and Space Weather models. The background estimation of the 3C50 model~\citep{remillard2022empirical} is plotted in the left panel of Figure~\ref{NICER_observation} in comparison with the total NICER energy spectrum. The instrumental responses include the effective area curve of the optics system (Ancillary Response Files, ARF) and the mapping from incident X-ray energy to the registered channel of the detection system (Redistribution Matrix Files, RMF), which will be used in the later-on pulse profile modeling. 

\begin{figure}[h]
\begin{center}
\includegraphics[width=0.8\linewidth]{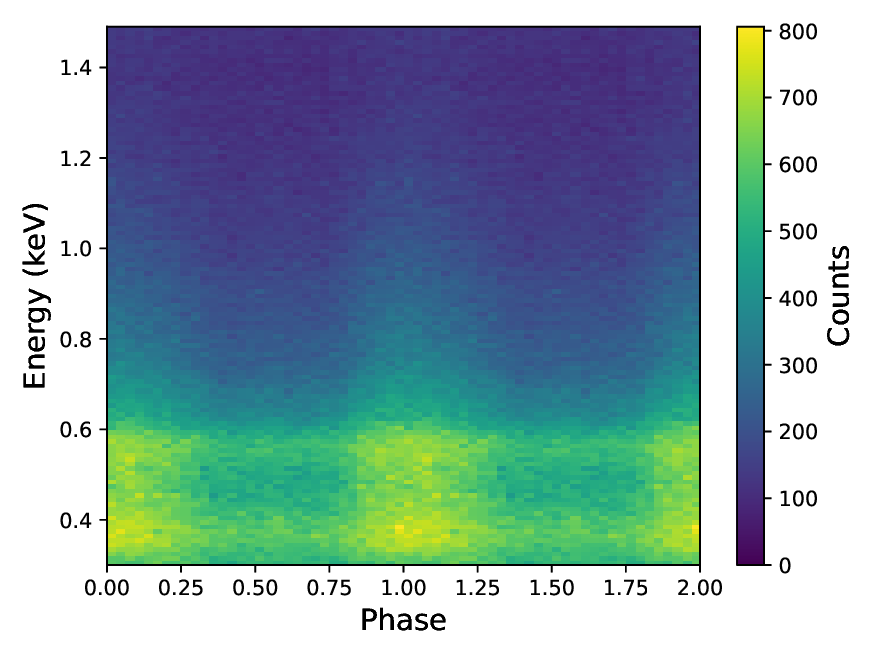}
\end{center}
\caption{Energy-dependent pulse profiles of PSR J1231--1411 in the range between 0.3 and 1.5~keV with 32 phase bins. Two rotational cycles are plotted for clarity. }
\label{NICER_observation2}
\end{figure}

The pulsar rotation phase of each photon is calculated by event folding~\citep{zheng2024new} using the timing model in Table~1 of \cite{ray2019discovery}. The timing residuals are then analyzed using tempo2~\citep{hobbs2006tempo2} based on the extracted average time of arrival and the timing model. The residuals are clustered around zero with only white noise remaining, and the root mean square is less than 10~$\mu$s, which demonstrates a proper timing model. The energy-integrated pulse profile in the range between 0.3 and 1.5~keV is plotted in the right panel of Figure~\ref{NICER_observation}. It is characterized by a prominent primary peak and a weak inter-pulse at $\approx$~0.55 rotation phase relative to the primary peak. The energy-dependent pulse profiles, a.k.a.\ phase-resolved energy spectra, are plotted in Figure~\ref{NICER_observation2}. These are the data to be fitted in the later-on pulse profile modeling.

\subsection{XMM-Newton}
\begin{figure}[h]
\begin{center}
\begin{minipage}{0.55\linewidth}
\includegraphics[width=1.0\linewidth]{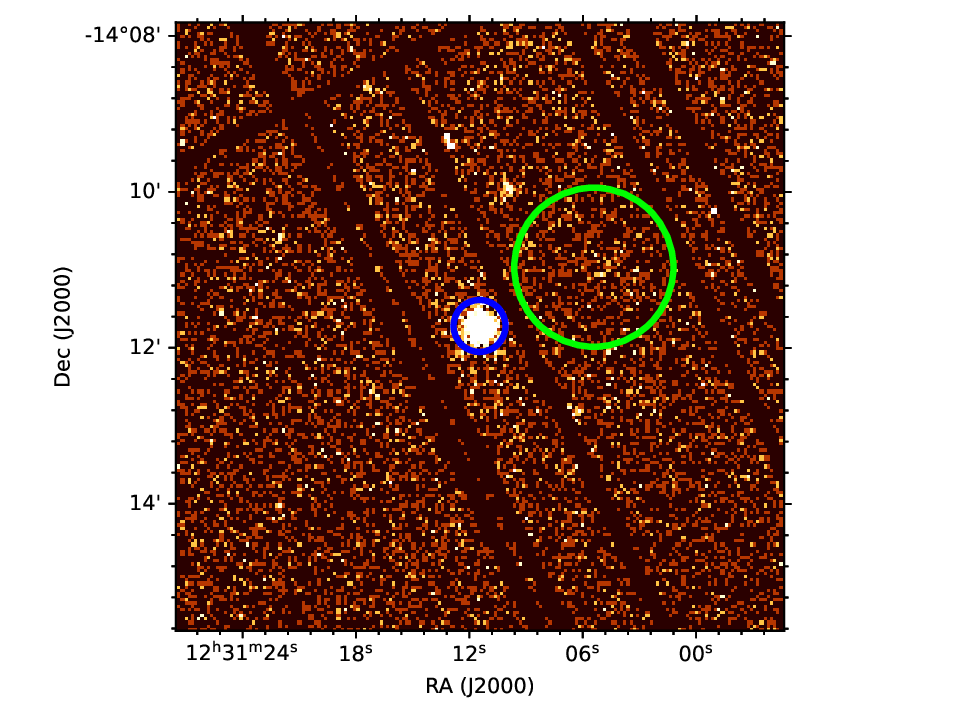}
\end{minipage}
\begin{minipage}{0.40\linewidth}
\includegraphics[width=1.0\linewidth]{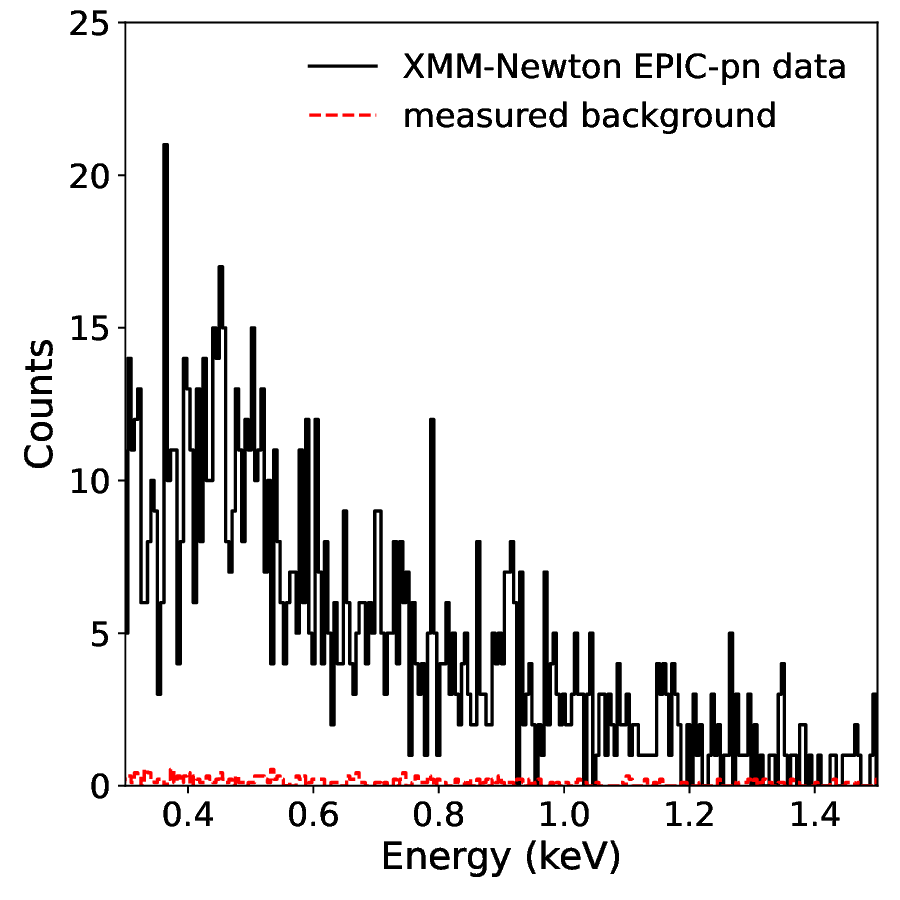}
\end{minipage}
\end{center}
\caption{Zoomed in image of PSR J1231--1411 with EPIC-pn filtered events (left panel). The source region is denoted in the blue circle. The background region is denoted in the green circle. Total XMM-Newton energy spectrum of PSR J1231--1411 and the measured background (right panel). }
\label{XMM_observation}
\end{figure}

PSR J1231--1411 was in the field of view of the XMM-Newton~\citep{jansen2001xmm} observation on July 15, 2009, with a duration of 29.8~ks (ObsID 0605470201){\footnote{https://www.cosmos.esa.int/web/xmm-newton/xsa}}. The observation data of EPIC-pn~\citep{struder2001european} is used for the later-on pulse profile modeling because it has a high source-to-background ratio in the energy spectrum. Since the timing resolution of the extended full-frame imaging mode prohibits the phase folding of PSR J1231--1411, only the phase-averaged spectral data of EPIC-pn will be used. EPIC-MOS1 and EPIC-MOS2 are not used in this work due to lower statistics than that of EPIC-pn. The data processing uses the XMM-Newton Science Analysis System (SAS version 18.0.0 and calibration database Update 2021-12-09). The observation data files are first resolved to event files through \textit{epproc}. To reduce the impact of background flares, an additional selection criteria of TIME~$>$~364045000 is used in \textit{evselect} along with the standard filtering expressions PATTERN~$\leq$~4, PI~$\in$~[200,15000], \#XMMEA\_EP, and FLAG==0. The final effective exposure is reduced to 13.43~ks after data filtering. The source spectrum is extracted within a 20-arcsecond radius circle centered at R.A.(J2000)~=~187.79768$^{\circ}$, Dec.(J2000)~=~-14.19546$^{\circ}$ (see the left panel of Figure~\ref{XMM_observation}). The scaling factor is calculated as 0.1075 (BACKSCAL) to correct the background estimation. Corresponding total energy spectrum and background estimation are plotted in the right panel of Figure~\ref{XMM_observation}. The ARF and RMF products of EPIC-pn are generated using the \textit{rmfgen} and \textit{arfgen} tools in SAS.

\section{Result and Discussion}
\label{sec4}
The numerical algorithm and the parameter inference process are validated in Section~\ref{sec3}. They are applied to the extracted observation data of PSR J1231--1411 (Section~\ref{sec2}) with the same configurations and settings except for the prior information and geometric model of the X-ray emitting region. The selection of geometric models follows the procedure in \cite{riley2019nicer}, which starts from the simplest model of two single-temperature circular spots to more complex models concerning the geometric shape and temperature distribution. A few geometric models are tested until a proper fit to the data is achieved. The preferred geometric model includes one single-temperature elongated region and one single-temperature crescent region. The definition of elongated hot spots inherits from the work of \cite{miller2019psr} to explain the data of PSR J0030+0451. It provides an alternative hot spot geometry possible for PSR J1231--1411 to explain simultaneously the observation data of NICER and XMM-Newton, complementary to more complex hot spot geometry of two dual-temperature overlapping regions in \cite{salmi2024nicer}. 

\subsection{\textbf{ST-U}}
The Single-Temperature-Unshared (\textbf{ST-U}) model is usually used first to fit the data, motivated by the physics of magnetospheric return currents with a dipole magnetic field~\citep{harding2001pulsar}. It is also relatively simple and computationally inexpensive with a limited number of model parameters. 

Inspired by the previous work~\citep{salmi2024nicer}, the \textbf{ST-U} model in this section defines the uniform priors on the radius in a range of [8,16]~km, distance to Earth [100,700]~pc, hydrogen column density [0,5]~$\times$10$^{20}$ cm$^{-2}$, cosine of the center colatitude [-1,1], angular radius [0,90]~deg, effective temperature [0.011,0.3]~keV, and center phase of the hot spot [a,a+0.2] (a and 0.2 are profile-dependent numbers). According to \cite{salmi2024nicer} and the reference therein (concerning the preliminary Shapiro delay measurement), a Gaussian prior PDF is defined on the mass and cosine of view inclination. The mean of the mass prior PDF is 1~\(\textup{M}_\odot\), and the standard deviation is 0.93~\(\textup{M}_\odot\), which is not a tight constraint. The Gaussian distribution is then truncated between 1 and 2~\(\textup{M}_\odot\), which is different from that of \cite{salmi2024nicer} (between 1 and 3~\(\textup{M}_\odot\)). Based on their inferred masses and the mass distribution of millisecond pulsars with a bimodal distribution~\citep{antoniadis2016millisecond}, PSR J1231--1411 is more likely to be a low-mass neutron star. The mean $\mu$ and standard deviation $\sigma$ of the cosine of view inclination are polynomials as a function of mass~\citep{salmi2024nicer},
\begin{equation}
\begin{aligned}
\mu_{\cos(i)}(M) = &-0.00835942\Big(\frac{M}{\textup{M}_\odot}\Big)^2 + 0.10890304\Big(\frac{M}{\textup{M}_\odot}\Big)\\
& + 0.01118702\,,
\end{aligned}
\end{equation} 
\begin{equation}
\begin{aligned}
\sigma_{\cos(i)}(M) = &-0.00097777\Big(\frac{M}{\textup{M}_\odot}\Big)^2 + 0.01013241\Big(\frac{M}{\textup{M}_\odot}\Big)\\
& + 0.01509739\,.
\end{aligned}
\end{equation} 
A Gaussian prior PDF is also defined on the energy-independent effective area scaling factor $\alpha$ ($\mu = 1$, $\sigma = 0.05$) for NICER and XMM-Newton, respectively. 

\begin{figure}[h]
\begin{center}
\begin{minipage}{0.49\linewidth}
\includegraphics[width=1.0\linewidth]{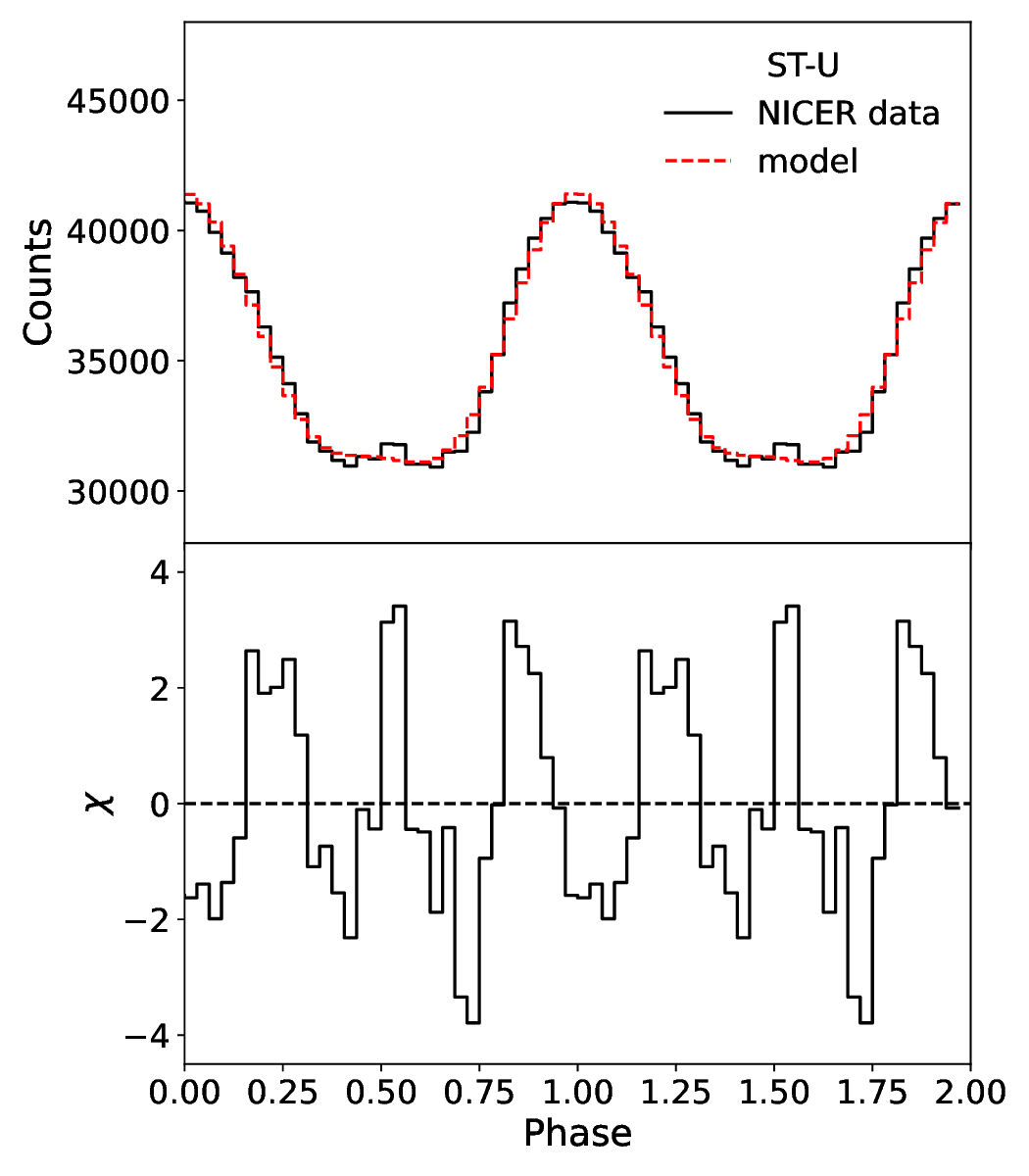}
\end{minipage}
\begin{minipage}{0.49\linewidth}
\includegraphics[width=1.0\linewidth]{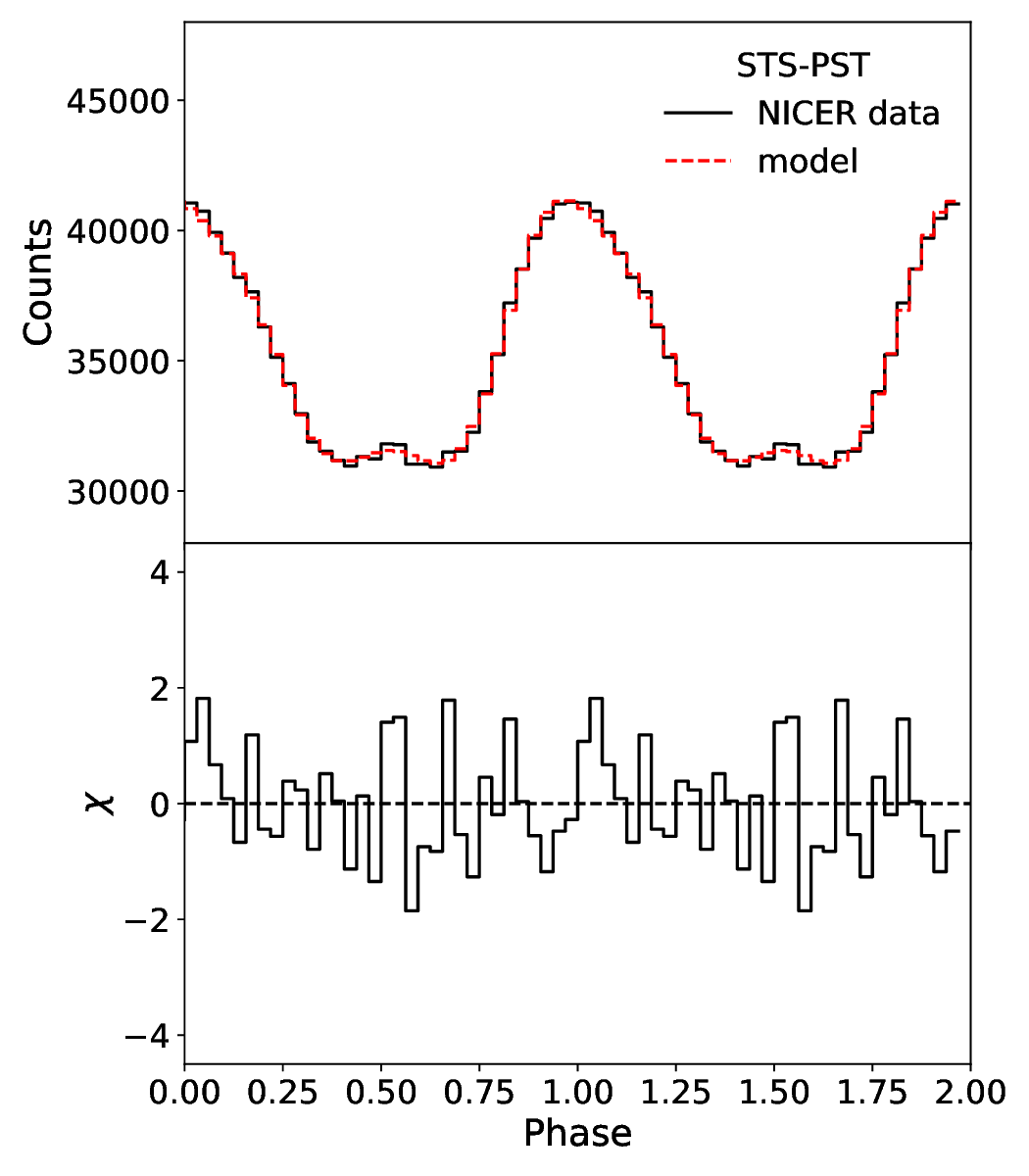}
\end{minipage}
\end{center}
\caption{Comparison between the best-fit energy-integrated pulse profile and the observation data for the \textbf{ST-U} (left panel) and \textbf{STS-PST} (right panel) models, respectively. The energy range of integration is from 0.3 to 1.5~keV. }
\label{comparison1}
\end{figure}

The inferred model parameters with 68\% credible intervals are summarized in Table~\ref{parameter_estimation}. The detailed 1-D and 2-D marginalized posteriors of the model parameters are plotted in the appendix (Figure~\ref{appendix1}). The posterior distribution has multiple distinct modes. It indicates that the estimated mean value of some parameters cannot fully represent the complexity of the underlying distribution. Especially, the inferred radius is very close to the prior boundary, which usually suggests that the Bayesian inference probably does not converge in the specified prior range. A forward calculation is performed to find the best-fit parameters and the background using the generated posterior samples by MultiNest. The best-fit parameters with the maximum log-likelihood value are also summarized in Table~\ref{parameter_estimation}. It provides a one-way test about how well the model fits the data by inspecting the residual distribution and $\chi^2$ of the energy-resolved pulse profiles, energy-integrated pulse profile, and total energy spectrum~\citep{miller2019psr}. The comparison between the best-fit energy-integrated pulse profile and the observation data is plotted in the left panel of Figure~\ref{comparison1}. The zigzag pattern is seen in the discrepancies between the model and the data. Additionally, the weak inter-pulse is not well reproduced by the model. These indicate an apparent model deficiency of \textbf{ST-U} for PSR J1231--1411. 

\begin{table*}[ht]
\centering
\caption{Summary of the Bayesian parameter inference of PSR J1231--1411 using the \textbf{ST-U} and \textbf{STS-PST} geometric models. }
\resizebox{\linewidth}{!}{
\begin{tabular}{lccccccc}
\hline
			& 			&\multicolumn{3}{c}{\textbf{ST-U}} &\multicolumn{3}{c}{\textbf{STS-PST}} \\ 
Parameter		& Description		&Prior &$\widehat{\textup{CI}}_{68\%}$ & Best-fit 
&Prior &$\widehat{\textup{CI}}_{68\%}$ & Best-fit\\
\hline
$F_0$ (Hz)		 		& spin frequency				
&271.45, fixed				& -	& -
&271.45, fixed				& -	& - \\

$M$ (\(\textup{M}_\odot\)) 		& gravitational mass			
&$M\sim$~Gaussian(1, 0.93)		& 1.36$_{-0.20}^{+0.18}$	&1.91
&$M\sim$~Gaussian(1, 0.93)		& 1.12$_{-0.07}^{+0.07}$	&1.08\\

&&
truncated between 1 and 2 \(\textup{M}_\odot\) &&&
truncated between 1 and 2 \(\textup{M}_\odot\)&&\\

$R_\textup{eq}$ (km) 				& equatorial circumferential radius 			
&$R_\textup{eq}\sim$~Uniform(8, 16)	& 9.89$_{-1.55}^{+0.98}$	&15.50
&$R_\textup{eq}\sim$~Uniform(8, 16)	& 9.91$_{-0.86}^{+0.88}$	&11.42\\

$\cos(i)$ 					& cosine of view inclination
&$\cos(i)\sim$~Gaussian($\mu(M)$, $\sigma(M)$)	& 0.15$_{-0.04}^{+0.04}$	&0.26
&$\cos(i)\sim$~Gaussian($\mu(M)$, $\sigma(M)$)	& 0.12$_{-0.02}^{+0.02}$	&0.10\\

$D$ (pc) 				& distance to Earth			
&$D\sim$~Uniform(100, 700)		& 573.75$_{-103.73}^{+97.22}$	&659.56
&$D\sim$~Uniform(100, 700)		& 621.72$_{-58.07}^{+57.97}$	&676.65\\

$N_\textup{H}$ (10$^{20}$ cm$^{-2}$)	& hydrogen column density 		
&$N_\textup{H}\sim$~Uniform(0, 5)	& 1.78$_{-0.46}^{+0.45}$	&1.63
&$N_\textup{H}\sim$~Uniform(0, 5)	& 2.31$_{-0.40}^{+0.40}$	&2.21\\

\hline
$\alpha_\textup{NICER}$ 			& NICER effective area scaling 	
&$\alpha_\textup{NICER}\sim$~Gaussian(1, 0.05)	& 0.98$_{-0.05}^{+0.05}$	&0.84
&$\alpha_\textup{NICER}\sim$~Gaussian(1, 0.05)	& 1.02$_{-0.04}^{+0.04}$	&1.02\\

$\alpha_\textup{XMM}$ 				& XMM-Newton effective area scaling 	
&$\alpha_\textup{XMM}\sim$~Gaussian(1, 0.05)	& 1.02$_{-0.05}^{+0.05}$	&1.15
&$\alpha_\textup{XMM}\sim$~Gaussian(1, 0.05)	& 0.97$_{-0.04}^{+0.04}$	&0.95\\

\hline
$\theta_\textup{p}$ (deg) 			& center colatitude of primary spot
&$\cos(\theta_\textup{p})\sim$~Uniform(-1, 1)	& 124.03$_{-11.09}^{+25.25}$	&23.59
&$\cos(\theta_\textup{p})\sim$~Uniform(-1, 1)	& 59.84$_{-14.57}^{+14.53}$	&39.69\\

$\Delta\theta_\textup{p}$ (deg) 			& angular radius of primary spot	
&$\Delta\theta_\textup{p}\sim$~Uniform(0.01, 90)	& 10.08$_{-2.33}^{+2.36}$	&8.69
&$\Delta\theta_\textup{p}\sim$~Uniform(0.01, 90)	& 5.24$_{-1.27}^{+1.31}$	&5.89\\

$kT_\textup{eff,p}$ (keV)			& effective temperature of primary spot
&$kT_\textup{eff,p}\sim$~Uniform(0.011, 0.3)	& 0.100$_{-0.004}^{+0.004}$	&0.092
&$kT_\textup{eff,p}\sim$~Uniform(0.011, 0.3)	& 0.084$_{-0.002}^{+0.002}$	&0.084\\

$\phi_\textup{p}$				& center phase of primary spot
&$\phi_\textup{p}\sim$~Uniform(a, a+0.2)	& a+0.154$_{-0.011}^{+0.012}$	&a+0.176
&$\phi_\textup{p}\sim$~Uniform(a, a+0.2)	& a+0.116$_{-0.005}^{+0.005}$	&a+0.113\\

$fs_\textup{p}$				& shape factor of primary spot
&-					& -			&-
&$fs_\textup{p}$ $\sim$~Uniform(1, 20)	& 6.59$_{-2.46}^{+2.60}$	&4.46\\

\hline
$\theta_\textup{s}$ (deg) 			& center colatitude of secondary spot
&$\cos(\theta_\textup{s})\sim$~Uniform(-1, 1)	& 96.84$_{-95.65}^{+82.20}$	&106.17
&$\cos(\theta_\textup{s})\sim$~Uniform(-1, 1)	& 61.55$_{-9.40}^{+9.53}$	&56.85\\

$\Delta\theta_\textup{s}$ (deg)	 			& angular radius of secondary spot	
&$\Delta\theta_\textup{s}\sim$~Uniform(0.01, 90)	& 27.07$_{-9.87}^{+10.72}$	&5.77
&$\Delta\theta_\textup{s}\sim$~Uniform(0.01, 90)	& 58.55$_{-6.21}^{+5.95}$	&52.50\\

$kT_\textup{eff,s}$ (keV)	 		& effective temperature of secondary spot
&$kT_\textup{eff,s}\sim$~Uniform(0.011, 0.3)	& 0.071$_{-0.006}^{+0.006}$	&0.066
&$kT_\textup{eff,s}\sim$~Uniform(0.011, 0.3)	& 0.077$_{-0.003}^{+0.003}$	&0.077\\

$\phi_\textup{s}$	 			& center phase of secondary spot
&$\phi_\textup{s}\sim$~Uniform(a+0.45, a+0.75)	& a+0.586$_{-0.066}^{+0.080}$	&a+0.712
&$\phi_\textup{s}\sim$~Uniform(a+0.45, a+0.75)	& a+0.569$_{-0.005}^{+0.005}$	&a+0.562\\

$f_\textup{s}$	 			& angular radius ratio of secondary spot
&-					& -	&-
&$f_\textup{s}$ $\sim$~Uniform(0.001, 1.999)	& 1.04$_{-0.02}^{+0.02}$	&1.04\\

$\varkappa_\textup{s}$ 			& angular separation of secondary spot
&-					& -	&-
&$\varkappa_\textup{s}$ $\sim$~Uniform(0, 1)	& 0.025$_{-0.018}^{+0.018}$	&0.022\\

$\varphi_\textup{s}$ (deg) 			& azimuthal offset of secondary spot
&-					& -	&-
&$\varphi_\textup{s}$ $\sim$~Uniform(0, 360)	& 175.36$_{-1.53}^{+1.50}$	&176.86\\

\hline
\end{tabular}
}
\label{parameter_estimation}
\end{table*}

\subsection{\textbf{STS-PST}}
Though the model deficiency of \textbf{ST-U} is found, its inferred parameters provide implications for possible hot spot geometries. In the \textbf{ST-U} model, the best-fit parameters show that the primary hot spot with a slightly higher temperature is very close to the north pole to be more visible in a rotation cycle. This is mainly due to the broad, prominent, and slightly asymmetric primary peak in the data, spanning over $\approx$~0.7 phases. The secondary hot spot is very close to the equator with a small radius to reproduce the weak inter-pulse in a much narrower phase range. To better fit the data within the prior range of mass [1,2]~\(\textup{M}_\odot\) and radius [8,16]~km, more complicated geometric models are needed in terms of the shape and temperature distribution. For example, the geometric model \textbf{PDT-U} (parsed as Protruding-Dual-Temperature-Unshared) was proposed in \cite{salmi2024nicer} to explain the data with two dual-temperature overlapping regions. As a supplement, a single-temperature elongated hot spot is used to fit the broad and prominent primary peak, and a single-temperature crescent hot spot is used for the narrow and weak inter-pulse in this work.

The definition of the elongated hot spot is based on \cite{miller2019psr} by introducing a colatitude-independent shape factor $fs$ to increase the longitudinal extent instead of a pure spherical cap, i.e.\ Single-Temperature with a Shape factor model (hereafter, \textbf{STS}). It should be noted that this definition produces oval spots in most cases, but it can also give ring-shaped hot spots when the longitudinal extent is larger than 2$\pi$ or fan-shaped hot spots when they are centered at the poles. The definition of the crescent hot spot \textbf{PST} (parsed as Protruding-Single-Temperature) follows \cite{riley2019nicer} with the angular radius ratio $f$, angular separation $\varkappa$, and azimuthal offset $\varphi$ between the superseding and ceding member. The uniform prior PDFs are defined on the shape factor $fs$ in a range of [1,20], angular radius ratio $f$ [0.001,1.999], angular separation $\varkappa$ [0,1], and azimuthal offset $\varphi$ [0,360]~deg. The prior PDFs remain unchanged for the other parameters in comparison to those defined in the \textbf{ST-U} model (see Table~\ref{parameter_estimation}).

\begin{figure}[h]
\begin{center}
\begin{minipage}{0.49\linewidth}
\includegraphics[width=1\linewidth]{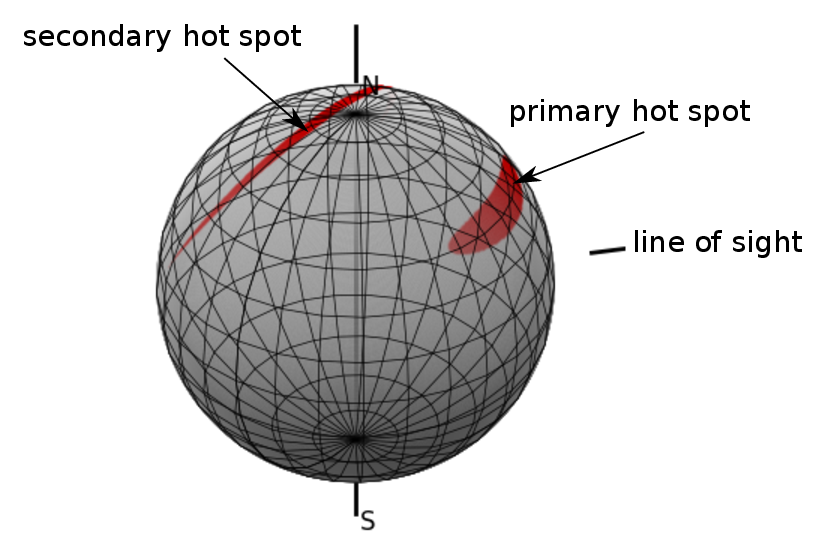}
\end{minipage}
\begin{minipage}{0.49\linewidth}
\includegraphics[width=1\linewidth]{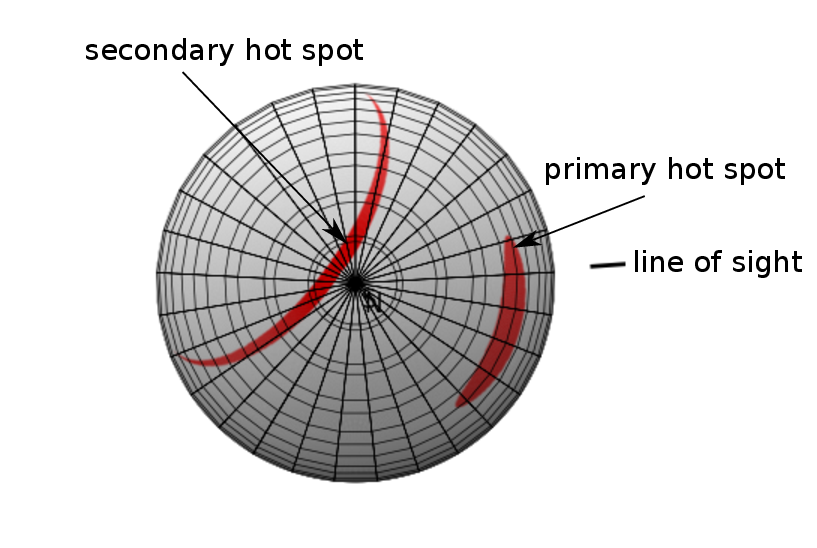}
\end{minipage}
\end{center}
\caption{Schematic view of the hot spot geometry according to the best-fit model. For the exact geometric parameters of the hot spots, refer to Table~\ref{parameter_estimation}.}
\label{geometry}
\end{figure}

\begin{figure}[h]
\begin{center}
\begin{minipage}{0.55\linewidth}
\includegraphics[width=1.0\linewidth]{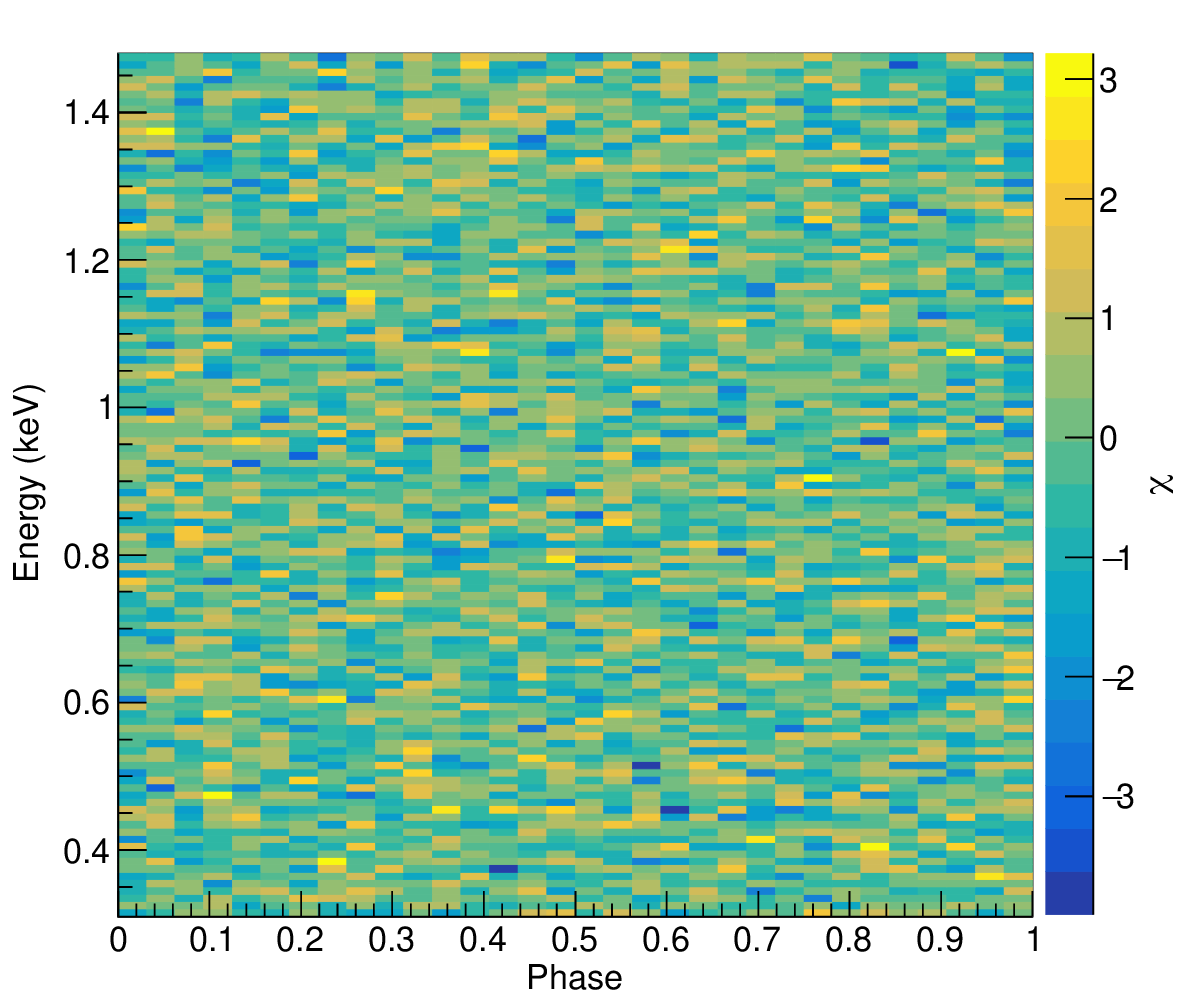}
\end{minipage}
\begin{minipage}{0.4\linewidth}
\includegraphics[width=1.0\linewidth]{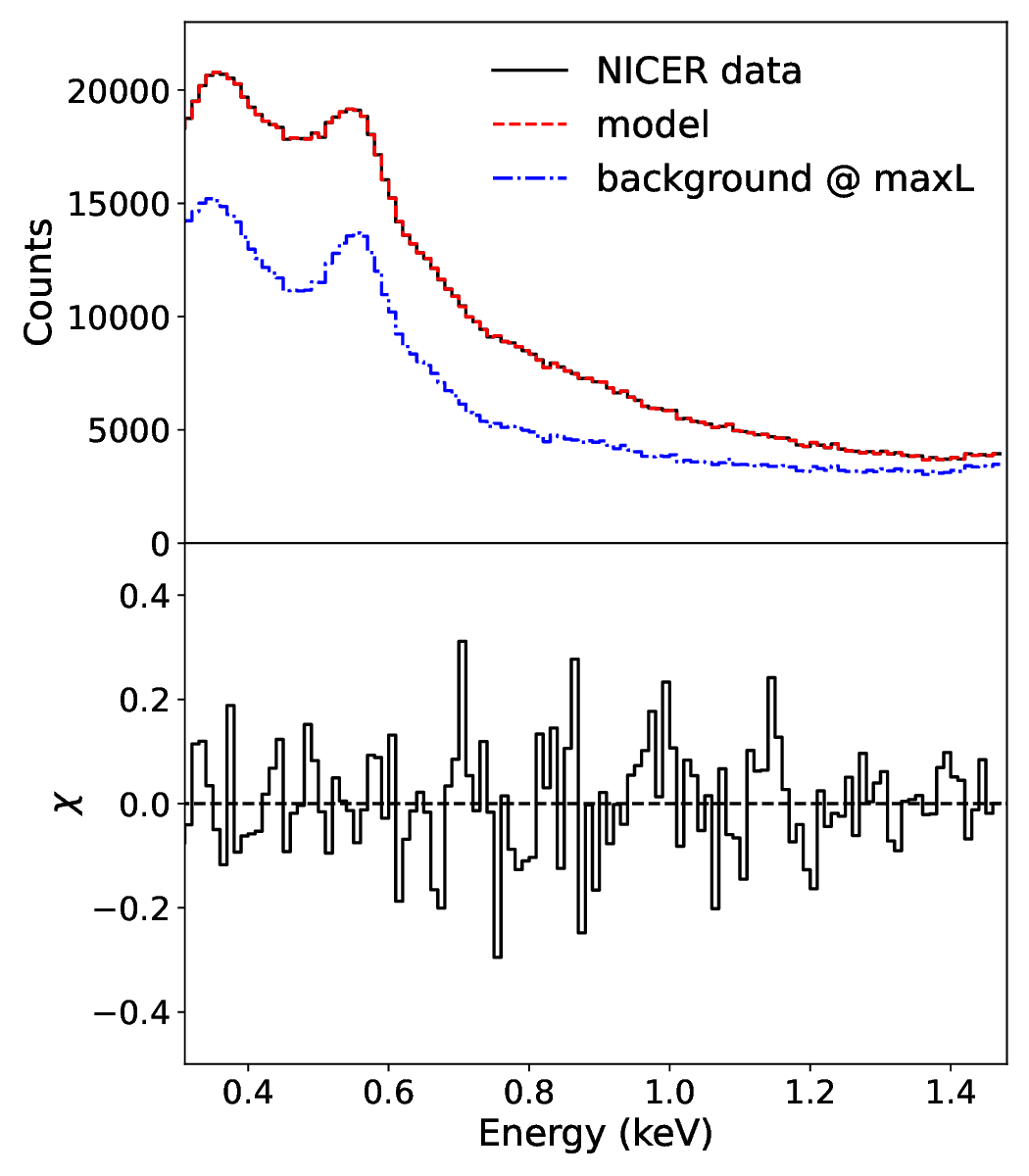}
\end{minipage}
\end{center}
\caption{Residuals of the energy-resolved pulse profiles between the best-fit model and the NICER data (left panel). Comparison of the total energy spectrum between the best-fit model and the NICER data (right panel). The background with the maximum log-likelihood value is also plotted.}
\label{residuals}
\end{figure}

\begin{figure}[h]
\begin{center}
\includegraphics[width=0.5\linewidth]{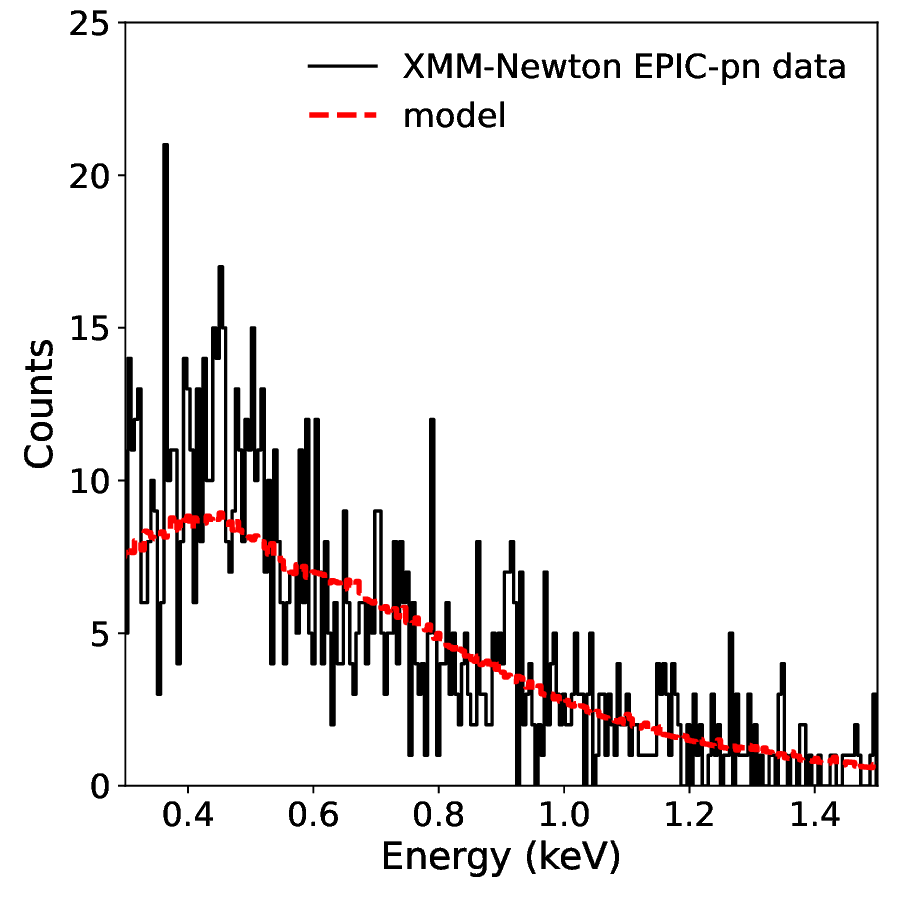}
\end{center}
\caption{Comparison of the total energy spectrum between the XMM-Newton data (EPIC-pn) and the best-fit model.}
\label{xmm_spectrum}
\end{figure}

The inferred model parameters with 68\% credible intervals are summarized in Table~\ref{parameter_estimation}. The detailed 1-D and 2-D marginalized posteriors of the model parameters are plotted in the appendix (Figure~\ref{appendix2}). With \textbf{STS-PST}, the inferred mass is $M~=~1.12~\pm~0.07$~\(\textup{M}_\odot\) and the inferred radius is $R_\textup{eq}~=~9.91_{-0.86}^{+0.88}$~km, which is consistent with previous analysis of NICER targets. A schematic view of the inferred hot spot geometry is plotted in Figure~\ref{geometry}, with both hot spots located in the northern hemisphere. The inferred geometry configuration is non-antipodal and suggests a complex magnetic field structure instead of a pure centered dipole magnetic field. To inspect the model adequacy of \textbf{STS-PST} for PSR J1231--1411, the comparison between the best-fit energy-integrated pulse profile and the observation data is plotted in the right panel of Figure~\ref{comparison1}. The discrepancies are decreased in comparison with those of the \textbf{ST-U} model, and the weak inter-pulse is reproduced by the \textbf{STS-PST} model. No apparent systematic structure is found in the residual distributions of the energy-resolved pulse profiles and the total energy spectrum (Figure~\ref{residuals}). Additionally, Figure~\ref{xmm_spectrum} shows that the best-fit model can reasonably explain the XMM-Newton data. 

The inferred hot spot geometry is similar to that of PSR J0030+0451, with both spots located in the same hemisphere~\citep{miller2019psr,riley2019nicer}. However, it is not consistent with that inferred by \textbf{PDT-U} in \cite{salmi2024nicer}. Another Bayesian inference is performed in this work to test the geometric model \textbf{STS-PST} by constraining the secondary hot spot in the southern hemisphere, i.e.\ $\cos(\theta_\textup{s})\sim$~Uniform(-1, 0). The inferred results from the partial parameter space are summarized in Table~\ref{parameter_estimation2} and Figure~\ref{appendix3}. The mass and radius are slightly increased but still consistent with those from the full parameter space calculation. The model calculation with the best-fit parameters also has a good agreement with the data, and no apparent systematic structure is found in the residual distributions. The Bayesian evidence and chi-square statistic of \textbf{ST-U} and \textbf{STS-PST} geometric models are summarized in Table~\ref{parameter_estimation3}. It depicts the model adequacy of \textbf{STS-PST}, which provides an alternative hot spot geometry possible for PSR J1231--1411 to explain simultaneously the observation data of NICER and XMM-Newton.

\begin{table}[ht]
\centering
\caption{Summary of the Bayesian parameter inference of PSR J1231--1411 using the \textbf{STS-PST} geometric models, where the secondary hot spot is restricted in the southern hemisphere. }
\begin{tabular}{lcc}
\hline
Parameter &$\widehat{\textup{CI}}_{68\%}$ & Best-fit  \\
\hline
$F_0$ (Hz)				& 271.45, fixed			& - \\
$M$ (\(\textup{M}_\odot\)) 		& 1.21$_{-0.11}^{+0.11}$	&1.16\\
$R_\textup{eq}$ (km) 			& 10.49$_{-0.82}^{+0.78}$	&11.25\\
$\cos(i)$ 				& 0.14$_{-0.02}^{+0.02}$	&0.16\\
$D$ (pc) 				& 632.52$_{-49.66}^{+48.05}$	&673.51\\
$N_\textup{H}$ (10$^{20}$ cm$^{-2}$)	& 2.38$_{-0.38}^{+0.38}$	&2.23\\
\hline
$\alpha_\textup{NICER}$ 		& 1.01$_{-0.04}^{+0.04}$	&0.97\\
$\alpha_\textup{XMM}$ 			& 0.98$_{-0.04}^{+0.04}$	&0.95\\
\hline
$\theta_\textup{p}$ (deg) 		& 80.33$_{-23.08}^{+31.74}$	&124.05\\
$\Delta\theta_\textup{p}$ (deg) 	& 3.73$_{-0.73}^{+0.74}$	&3.74\\
$kT_\textup{eff,p}$ (keV)		& 0.085$_{-0.003}^{+0.003}$	&0.087\\
$\phi_\textup{p}$			& a+0.114$_{-0.005}^{+0.006}$	&a+0.102\\
$fs_\textup{p}$				& 10.10$_{-2.90}^{+2.95}$	&10.15\\
\hline
$\theta_\textup{s}$ (deg) 		& 143.68$_{-6.38}^{+6.50}$	&153.50\\
$\Delta\theta_\textup{s}$ (deg)	 	& 43.45$_{-4.49}^{+4.28}$	&36.62\\
$kT_\textup{eff,s}$ (keV)	 	& 0.083$_{-0.003}^{+0.003}$	&0.088\\
$\phi_\textup{s}$	 		& a+0.551$_{-0.005}^{+0.006}$	&a+0.541\\
$f_\textup{s}$	 			& 1.08$_{-0.04}^{+0.04}$	&1.04\\
$\varkappa_\textup{s}$ 			& 0.091$_{-0.052}^{+0.052}$	&0.25\\
$\varphi_\textup{s}$ (deg) 		& 4.65$_{-1.24}^{+1.18}$	&5.34\\
\hline
\end{tabular}
\label{parameter_estimation2}
\end{table}

\begin{table}[ht]
\centering
\caption{Summary of the Bayesian evidence and chi-square statistic of \textbf{ST-U} and \textbf{STS-PST} geometric models. }
\begin{tabular}{lcc}
\hline
Model 						& Log-evidence 		& $\chi^2$/d.o.f.  \\
\hline
\textbf{ST-U}					& -15670.24		&3.88\\
\textbf{STS-PST} full parameter space		& -15621.85		&0.96\\
\textbf{STS-PST} partial parameter space	& -15623.01		&0.95\\
\hline
\end{tabular}
\label{parameter_estimation3}
\end{table}

\section{Conclusions}
\label{sec5}
A numerical algorithm for pulse profile modeling is implemented in this work based on the spherical star Schwarzschild-spacetime and Doppler approximation. The accuracy of the code is validated by re-analyzing three rotation-powered millisecond pulsars, including PSR J0030+0451, PSR J0740+6620, and PSR J0437--4715. The inferred mass--radius relations and X-ray emitting region properties are consistent with other works, which allows the application of the numerical algorithm to PSR J1231--1411. Additionally, it can be modified and used to generate input data for performing end-to-end observation simulations of future NICER-like X-ray missions, e.g. the enhanced X-ray Timing and Polarimetry (eXTP)~\citep{zhang2019enhanced,watts2019dense}.

According to \cite{salmi2024nicer}, the Bayesian parameter inference in this work uses similar prior information, including the mass, radius, inclination, distance, hydrogen column density, and instrumental scaling factor. On the other hand, an alternative geometry configuration of X-ray emitting regions, i.e.\ one single-temperature elongated region and one single-temperature crescent region, is adopted to fit the observation data of NICER and XMM-Newton jointly. The model with the best-fit parameters can explain the data with acceptable residuals. The inferred gravitational mass is $M~=~1.12~\pm~0.07$~\(\textup{M}_\odot\) and the inferred equatorial radius is $R_\textup{eq}~=~9.91_{-0.86}^{+0.88}$~km. 
The inferred mass is slightly higher, and the inferred radius is smaller than that derived by \citet{salmi2024nicer} with two dual-temperature overlapping regions and an informative radius prior. Both analyses can properly fit the bolometric and energy-resolved pulse profiles and recover the weak interpulse. It is probably due to less tight constraints on the mass and inclination prior from the preliminary radio pulsar timing analysis, uncertainties of the distance measurements, and a weak inter-pulse in the pulse profile.
The inferred geometry configuration is non-antipodal and suggests a complex magnetic field of PSR J1231--1411 instead of a pure centered dipole magnetic field. 

\begin{acknowledgments}
This research utilized data and software from the High Energy Astrophysics Science Archive Research Center (HEASARC), provided by NASA's Goddard Space Flight Center. This research also utilized data based on observations obtained with XMM-Newton, an ESA science mission with instruments and contributions directly funded by ESA Member States and NASA. This work is supported by the National Key R\&D Program of China (2021YFA0718500) from the Minister of Science and Technology of China (MOST). The authors thank supports from the National Natural Science Foundation of China (Grant Nos. 12373051, 12333007, and 12273028) and the International Partnership Program of Chinese Academy of Sciences (Grant No. 113111KYSB20190020). We thank Hong LI and the project of Ali CMB Polarization Telescope (AliCPT) for providing the computation resources. 
\end{acknowledgments}

\bibliography{mybib}{}
\bibliographystyle{aasjournal}

\appendix

\begin{figure}[h]
\begin{center}
\includegraphics[width=0.88\linewidth]{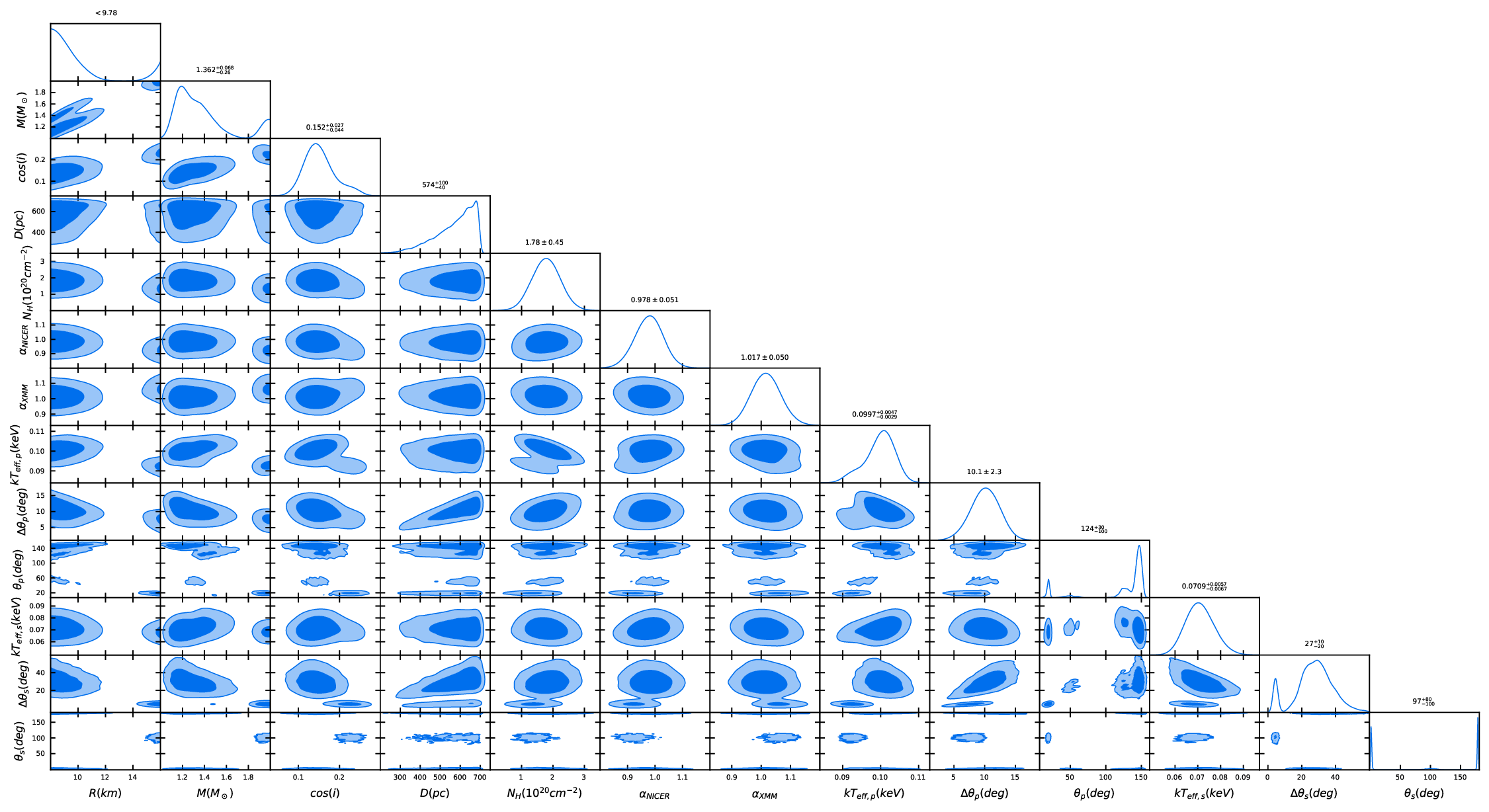}
\end{center}
\caption{1-D and 2-D posterior PDFs of the model parameters in \textbf{ST-U} for PSR J1231--1411.}
\label{appendix1}
\end{figure}

\begin{figure}[h]
\begin{center}
\includegraphics[width=0.88\linewidth]{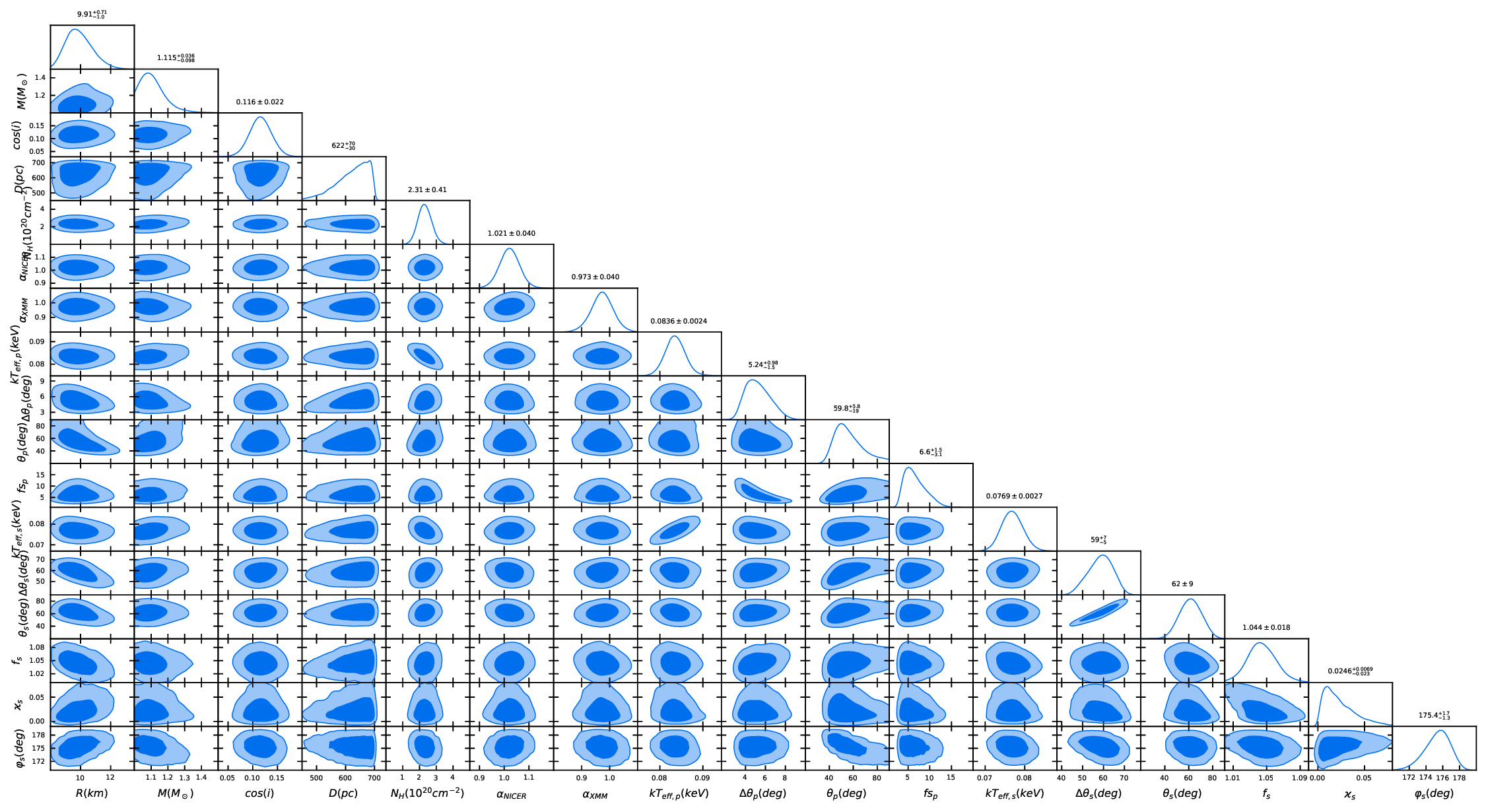}
\end{center}
\caption{1-D and 2-D posterior PDFs of the model parameters in \textbf{STS-PST} for PSR J1231--1411.}
\label{appendix2}
\end{figure}

\begin{figure}[h]
\begin{center}
\includegraphics[width=0.88\linewidth]{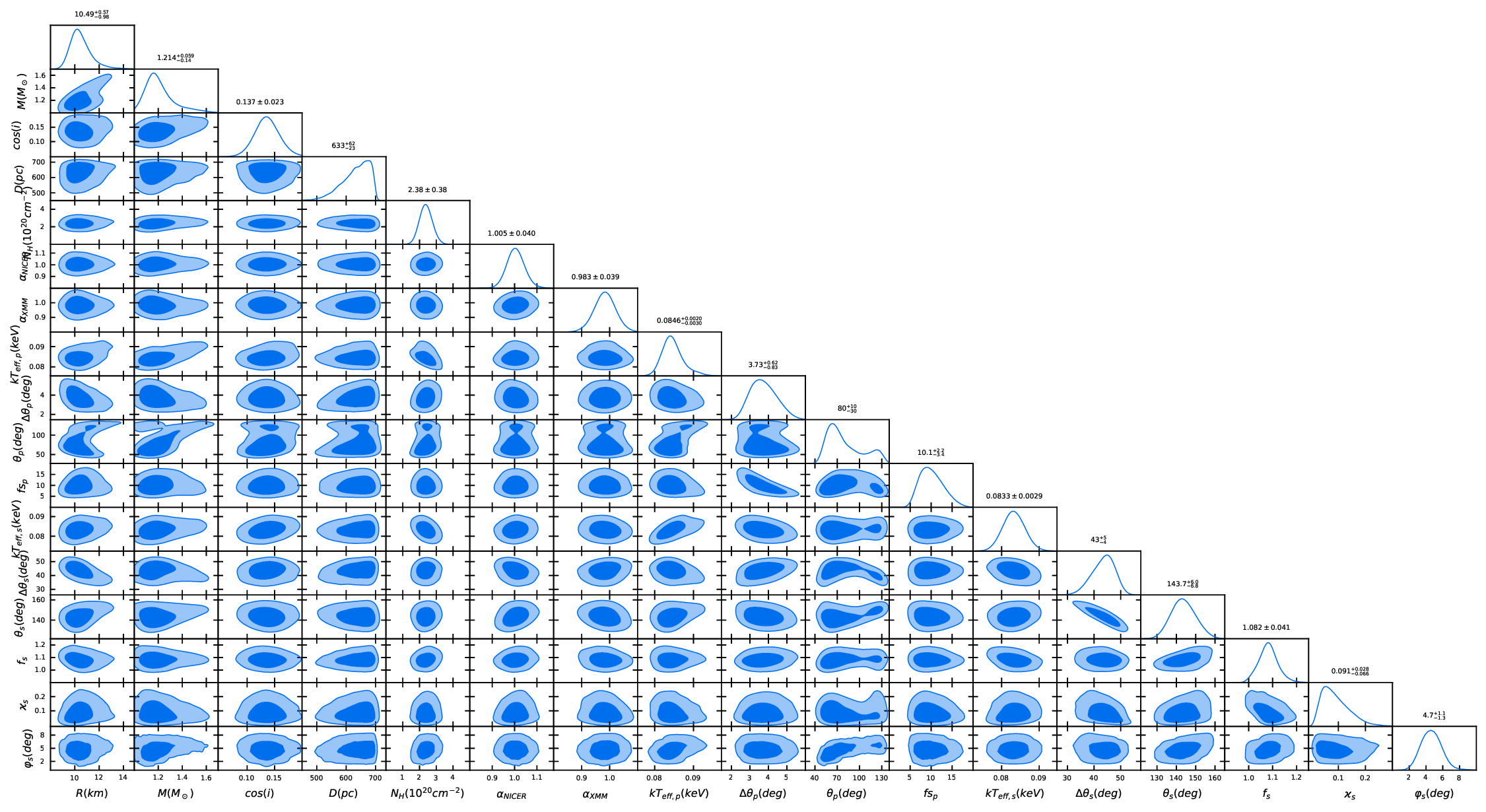}
\end{center}
\caption{1-D and 2-D posterior PDFs of the model parameters in \textbf{STS-PST} for PSR J1231--1411, where the secondary hot spot is restricted in the southern hemisphere.}
\label{appendix3}
\end{figure}

\end{document}